\documentclass{aa}    

\usepackage{amsmath}
\usepackage{amssymb}
\usepackage{epsfig}
\usepackage{graphicx}
\usepackage{natbib}
\usepackage{multirow}

  \newcommand{\Rsolar}{\mbox{\,$R_{\odot}$}}        

  \newcommand{\kms}{\,\rm km^2 \rm s^{-1}}
  \newcommand{\dgr}{^{\circ}}

\begin{document}


\title{A necessary extension of the surface flux transport model} 
\titlerunning{A necessary extension of the surface flux transport model}

\author{I. Baumann\thanks{Present address: Observatoire Royal de Belgique,
 Avenue Circulaire 3, 1180 Bruxelles, Belgium}
 \and D. Schmitt \and M. Sch\"ussler}
\institute{Max-Planck-Institut f\"ur Sonnensystemforschung, 
           37191 Katlenburg-Lindau, Germany}

\abstract{Customary two-dimensional flux transport models for the
evolution of the magnetic field at the solar surface do not account for
the radial structure and the volume diffusion of the magnetic
field. When considering the long-term evolution of magnetic flux, this
omission can lead to an unrealistic long-term memory of the system and
to the suppression of polar field reversals. In order to avoid such
effects, we propose an extension of the flux transport model by a linear
decay term derived consistently on the basis of the eigenmodes of the
diffusion operator in a spherical shell. A decay rate for each
eigenmode of the system is determined and applied to the corresponding
surface part of the mode evolved in the flux transport model. The value
of the volume diffusivity associated with this decay term can be
estimated to be in the range 50--100$\,\kms$ by considering the
reversals of the polar fields in comparison of flux transport
simulations with observations. We show that the decay term prohibits a
secular drift of the polar field in the case of cycles of varying
strength, like those exhibited by the historical sunspot record.}

\date{\today}

\maketitle

\keywords{Sun: magnetic fields - 
          Sun: photosphere - Magnetohydrodynamics (MHD) }

\section{Introduction}
\label{sec:intro}

Flux transport models describe the evolution of the flux distribution at
the solar surface as a result of the emergence of bipolar magnetic
regions and the transport of the corresponding radial magnetic flux by
the horizontal flows due to convection, differential rotation and
meridional circulation \citep[e.g.,][]{Leighton64, DeVore84,
WNS89Science, BCP98, sch01jan, Mackay1, Durrant:etal:2004,
Baumann2004}. When applying the model to multiple solar cycles,
\citet{sch02oct} and \citet{WLS2002} noticed that the secular evolution
resulting from the simulations is not in agreement with the
observations: activity cycles of varying strength lead to a drift of the
polar fields and even to the disappearance of polar reversals.

The conceptual deficiency in the conventional flux transport models that
leads to this disagreement with the observations arises from ignoring
the vectorial nature and the radial structure of the magnetic field and,
particularly, from omitting the part of the diffusion operator depending
on the radial coordinate \citep[see also][]{Wilson:etal:1990,
DikChoud94}. This leads to an unwanted long-term memory of the surface
flux since the decay time of a dipolar surface field increases strongly
in the presence of a poleward meridional flow. 

In order to avoid such unrealistic effects, \citet{WLS2002} made the
assumption that the speed of the meridional flow is higher during
more active cycles, while \citet{sch02oct} introduced an ad-hoc local
decay term, calling upon the `sea-serpent' flux emergence process
combined with coronal field line reconnection as proposed by
\citet{Spruit:etal:1987}. Alternatively, \citet[][and earlier references
therein]{DikChoud99} have considered a consistent MHD model for the
evolution of an axisymmetric field in the meridional plane of the
Sun. Unfortunately, our understanding of the solar dynamo and the
available computational resources are still insufficient to carry out
such simulations in the three-dimensional case. We therefore suggest to
extend the flux transport model by a modal version of the decay term
proposed by \citet{sch02oct}. We derive this term by considering the
decay of the eigenmodes of the (volume) diffusion operator in a
spherical shell. This leads to a consistent decay rate for each
eigenmode, which can simply be incorporated in a code based upon
expansion of the magnetic field into surface harmonics. The volume
diffusion coefficient remains as a free parameter, which can be
estimated by comparison of simulation results with observational
constraints (like polar field reversals)

The outline of this paper is as follows. In Sect.~\ref{sec:decaymodes}
we determine the decay modes in a spherical shell for appropriate
boundary conditions. The results are used in Sect.~\ref{sec:extension}
to define the decay term extending the surface flux transport model. The
value of the volume diffusivity (free parameter in the deacy term) is
estimated by considering the evolution of the polar field over activity
cycles of varying strength in Sect.~\ref{sec:cycle_var} and by comparing
the times of polar reversals with observational data in
Sect.~\ref{sec:reversals}.  We summarize our results in
Sect.~\ref{sec:conclusion}.

\section{Decay modes of a poloidal field in a spherical shell}
\label{sec:decaymodes}

In the abscence of systematic flows, the time evolution of a magnetic
field is described by the diffusion equation,
\begin{eqnarray}
  \frac{\partial\bf{B}}{\partial t} =  - \eta\,\nabla\times 
  (\nabla\times{\bf B}) \, ,
\label{DiffusionEquation}
\end{eqnarray}
where $\eta$ is the (constant) magnetic diffusivity.  In spherical
geometry, the magnetic field vector can be written as the sum of a
poloidal and a toroidal part, the latter having no radial component.
Since the flux transport model assumes a purely radial magnetic field at
the solar surface, it is sufficient to consider the decay of a poloidal
magnetic field in a spherical shell (representing the solar convection
zone), setting the non-radial components to zero at the outer boundary
(the solar surface).  The evolution of this field is described by
Eq.~(\ref{DiffusionEquation}), where $\bf{B}$ now represents a poloidal
field. We introduce a spherical polar coordinate system with coordinates
$(r,\theta,\phi)$, whose origin is located in the center of a sphere of
radius $R_{\sun}$.  The poloidal magnetic field can be represented by a
scalar function, $S(r,\theta,\phi,t)$, as
\begin{eqnarray}
{\bf B} = -\nabla \times ({\bf r} \times \nabla S) = -{\bf r} \,\triangle\; S + \nabla \frac{\partial}{\partial r} (rS) \, ,
\label{Bpol1}
\end{eqnarray}
where $\triangle$ is the Laplace operator in spherical coordinates and
{\bf r} is the radius vector \citep{BullardGellman1954,KrauseRaedler}
. Inserting this field representation into Eq.~(\ref{DiffusionEquation})
we obtain
\begin{eqnarray}
  -\nabla \times\left( {\bf r} \times \nabla \; \frac{\partial S}{\partial t} \right) 
&=&  \eta\; \nabla \times\nabla \times\nabla \times \left( {\bf r} \times \nabla S \right) \nonumber \\
&=& \eta\;\nabla \times  \nabla \times \left( {\bf r} \times \triangle  S \right)  \nonumber \\
[1ex]
&=& -\eta\;\nabla \times  \left( {\bf r} \times  \nabla \triangle S \right) \, ,
\label{DiffusionEquationPolS1}
\end{eqnarray}
where we have used $\nabla \times {\bf r} =0$. From Eq.~(\ref{DiffusionEquationPolS1}) it follows that
\begin{eqnarray}
 {\bf r} \times \nabla \left[ \eta \triangle S - \frac{\partial S}{\partial t} \right] = 0 \, .
\label{DiffusionEquationS2}
\end{eqnarray}
The vector potential, $-{\bf r}\times\nabla S$, is invariant under gauge transformations, so that $S$ can be choosen such that the normalization condition
\begin{eqnarray}
\int\limits_{-1}^{1} \int\limits_{0}^{2\pi} S \;d(\cos\theta) \; d\phi = 0
\label{NormS}
\end{eqnarray}
is fulfilled for each value of $r$. Denoting by $\Omega$ the angular part of
the spherical Laplace operator,
\begin{eqnarray}
\triangle S &=& \frac{1}{r^2} \bigg[ \frac{\partial}{\partial r}\left( r^2 \frac{\partial S}{\partial r}  \right) \nonumber\\[1ex]
&+& \frac{1}{\sin\theta}\frac{\partial}{\partial \theta} \left( \sin\theta  \frac{\partial S}{\partial \theta} \right) + \frac{1}{\sin\theta} \frac{\partial^2 S}{\partial \phi^2} \bigg] \nonumber \\[1ex]
&\equiv& \frac{1}{r^2} \bigg[ \frac{\partial}{\partial r}\left( r^2 \frac{\partial S}{\partial r}  \right) +  \Omega S \bigg]\, ,
\label{DefOmega}
\end{eqnarray}
we have
\begin{eqnarray}
\int\limits_{-1}^{1} \int\limits_{0}^{2\pi} \Omega S \;d(\cos\theta) \; d\phi = 0 \; .
\label{IntOmega}
\end{eqnarray}
With the normalization condition, Eq.~(\ref{NormS}), we then find
\begin{eqnarray}
\int\limits_{-1}^{1} \int\limits_{0}^{2\pi} \triangle S \;d(\cos\theta) \; d\phi = 0 \, .
\label{IntLaplace}
\end{eqnarray}
Using this result, we obtain from Eq.~(\ref{DiffusionEquationS2}) the
scalar diffusion equation 
\begin{eqnarray}
\eta \triangle S - \frac {\partial S}{\partial t} = 0 \, .
\label{DiffusionEquationSInn}
\end{eqnarray}
We search for solutions of this equation within a spherical shell,
$r_{\rm b}\leq r \leq R_{\sun} $, which are uniquely determined by
specifying boundary conditions at the inner and outer boundaries. In
accordance with the flux transport model, we assume the magnetic field
to be radial at the surface. This leads to the boundary condition
\begin{eqnarray}
  \frac{\partial (rS)}{\partial r} = 0 \quad \quad \mbox{at}  
        \quad r=R_{\odot}\, .
\label{BC_oben}
\end{eqnarray}
\begin{table*}
  \caption{Decay times $\tau_{ln} = 1/(\eta\, k_{ln}^2)$ (in years) for
  a volume diffusion coefficient of $\eta=100\,$km$^2\,$s$^{-1}$. The
  modes are characterized by $l$, the number of node circles on
  spherical surfaces, and $n$, the number of nodes in the radial
  direction.}
    \centering
  \begin{tabular}{cc|ccccccccc}
    \hline\hline
    & & \multicolumn{9}{c}{$l$}\\
    & & 1 & 2 & 3 & 4 & 5 & 6 & 7 & 8 & 9 \\
    \hline 
    \multirow{5}{*}{$n$} & 0  &     5.15  &     4.42  &     3.65  &     2.97  &     2.40  &     1.96  &     1.61  &     1.34  &     1.13\\
    &  1  &     0.62  &     0.60  &     0.58  &     0.56  &     0.53  &     0.50  &     0.47  &     0.44  &     0.41\\
    &  2  &     0.22  &     0.22  &     0.22  &     0.22  &     0.21  &     0.21  &     0.20  &     0.20  &     0.19\\
    &  3  &     0.11  &     0.11  &     0.11  &     0.11  &     0.11  &     0.11  &     0.11  &     0.11  &     0.10\\
    &  4  &     0.07  &     0.07  &     0.07  &     0.07  &     0.07  &     0.07  &     0.07  &     0.07  &     0.07\\
    \hline
  \end{tabular}
  \label{table1} 
\end{table*}
The bottom of the convection zone borders on the radiative core, which
we represent by a field-free ideal conductor. Consequently, we require
as boundary condition at $r=r_{\rm b}$ that the radial component of the
magnetic field vanishes, which is equivalent to the condition
\begin{eqnarray}
  S = 0 \quad \quad  \mbox{at} \quad r=r_{\rm b}\, .
\label{BC_unten}
\end{eqnarray}
In the numerical calculations below we take  $r_{\rm b}=0.7R_{\sun}$, which
corresponds to the bottom of the solar convection zone.

The general solution of Eq.~(\ref{DiffusionEquationSInn}) can be
written as a decompostion into orthogonal decay modes \citep{Elsasser46},
\begin{eqnarray}
 S(r,\theta,\phi,t) = \sum_{n=0}^{\infty} \sum_{l=1}^{\infty}
\sum_{m=-l}^l R_{ln}(r) \; Y_{lm}(\theta,\phi) \; T_{ln}(t) \, ,
\label{SinSH}
\end{eqnarray}
where we have omitted the monopole term ($l=0$). The functions $Y_{lm}$
are the spherical surface harmonics. Separation of variables leads to 
an exponential time dependence, viz.
\begin{eqnarray}
T_{ln} (t) = \exp(-\eta\, k_{ln}^2 t) \, ,
\label{TimeDepPart}
\end{eqnarray}
where $1/\eta\, k_{ln}^2$ is the decay time of the mode characterized by
the wave numbers $l$ and $n$. For the 
spherical harmonics we have
\begin{eqnarray}
 \triangle \; Y_{lm} (\theta,\phi) =  -\frac{l\;(l+1)}{R^2_{\odot}}\;Y_{lm} (\theta,\phi) \, ,
\label{VecId6}
\end{eqnarray}
so that we obtain the following differential equation for the functions
$R_{ln}(r)$: 
\begin{eqnarray}
r^2\frac{{\rm d}^2 R_{ln}}{{\rm d}{r}^2} +2r\frac{{\rm d} R_{ln}}{{\rm d}{r}} + \bigg[ k_{ln}^2 {r}^2 - l(l+1)\bigg] R_{ln}  = 0 \, .
\label{Bessel}
\end{eqnarray}
Solutions are the  spherical Bessel functions of the first and
second kind, $j_l$ and $y_l$, respectively, so that
we have the general solution
\begin{eqnarray}
R_{ln} (r) = a_{ln}\; j_l(k_{ln}r) +  b_{ln} \; y_l(k_{ln}r)\, .
\label{Bessel3}
\end{eqnarray}
The linearity of the diffusion equation allows us to set the
coefficients $a_{ln}$ to unity 
without loss of generality. The lower boundary condition,
Eq.~(\ref{BC_unten}), is then used to determine
$b_{ln}=-j_l(k_{ln}r_{\rm b}) / y_l(k_{ln}r_{\rm b})$. Inserting this into
Eq.~(\ref{Bessel3}) and using the  upper
boundary condition, Eq.~(\ref{BC_oben}), leads to
\begin{eqnarray}
\lefteqn{l\;\Big[j_l(k_{ln}R_{\sun})\;y_l(k_{ln}r_{\rm b}) 
   - j_l(k_{ln}r_{\rm b})\;y_l(k_{ln}R_{\sun}) \Big] } \nonumber \\ 
[1ex]
\lefteqn{\qquad -k_{ln}R_{\sun} \Big[j_{l-1}(k_{ln}R_{\sun})\;y_l(k_{ln}r_{\rm b})} 
\nonumber \\ 
[1ex]
\lefteqn{\qquad\qquad\qquad - j_l(k_{ln}r_{\rm b})\;y_{l-1}(k_{ln}R_{\sun}) \Big]=0\; ,}
\label{Bessel4}
\end{eqnarray}
from which the eigenvalues $k_{ln}$ can be determined numerically. The
index $n$ is the number of nodes of the eigenfunction $R_{ln}(r)$ in
$r_{\rm b} < r < R_{\sun}$.  Note that the above construction shows that
the eigenvalues, $k_{ln}$, and, therefore, the temporal decay rates of
the eigenmodes do not depend on the azimuthal wave number, $m$, which
describes the $\phi$-dependence of the mode.

Table~\ref{table1} gives the decay times, $\tau_{ln} = 1/(\eta\,
k_{ln}^2)$, in years for various modes, assuming a volume diffusion
coefficient of $\eta=100\,$km$^2\,$s$^{-1}$. The dipole mode ($l=1$)
with $n=0$ has the longest decay time of about 5 years. While the decay
times decrease only slowly for the higher multipoles (increasing $l$),
the decay of the higher radial modes is much more rapid for the higher
radial modes, being faster by  a factor eight already for the mode
$(l=1, n=1)$. It is therefore justified to consider only the most slowly
decaying modes with $n=0$ in the decay term for the surface transport
model.  

\section{Extension of the surface flux transport model}
\label{sec:extension}

\begin{figure*}
\centering
\resizebox{0.9\hsize}{!}{\includegraphics[angle=270]{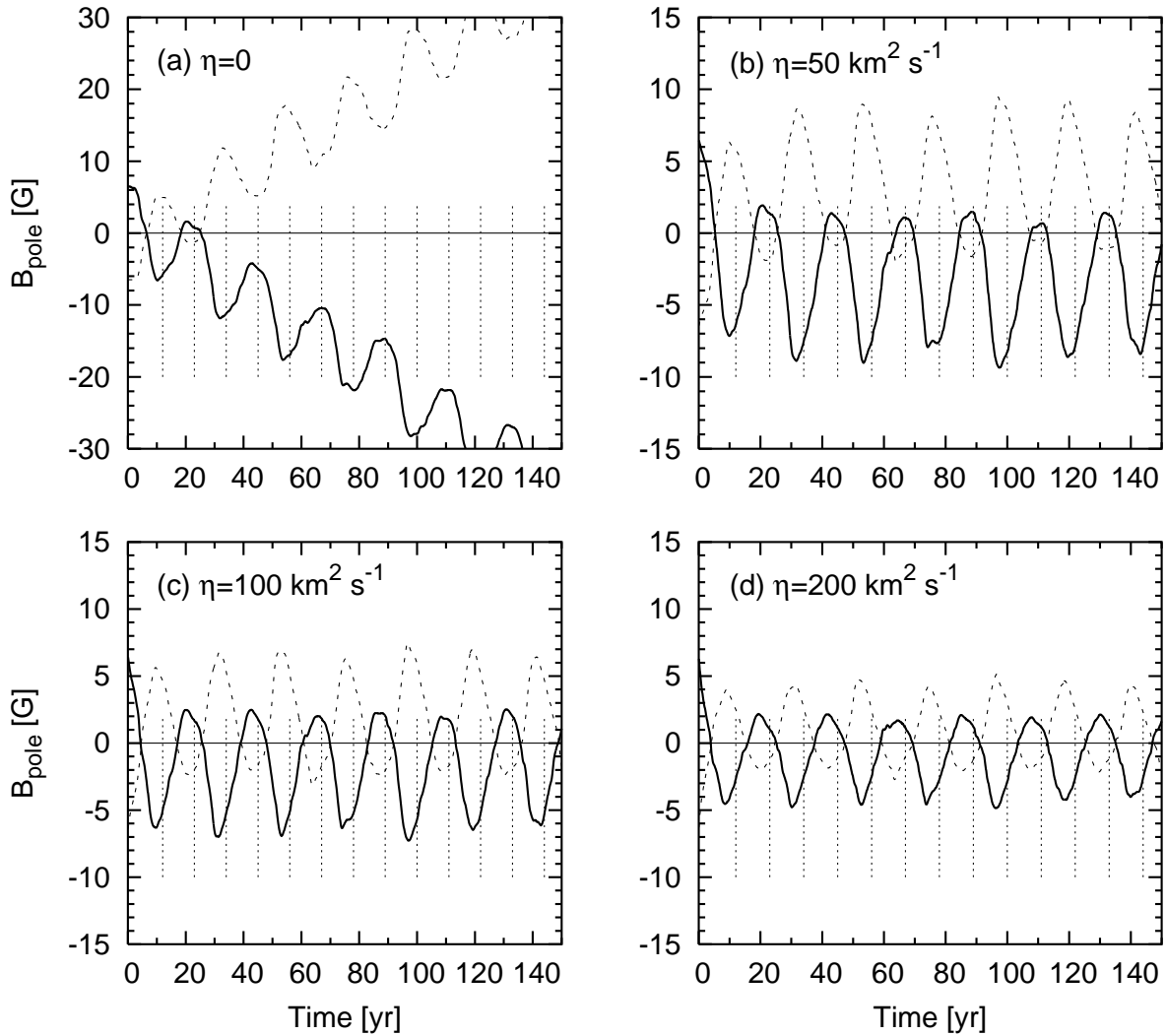}}
\caption{Evolution of the north (solid) and south polar fields (dashed)
in the case of a systematic variation of strength between even and odd
cycles for four values of $\eta$.  Finite values of $\eta$ lead to
stable oscillations of the polar fields while the field reversals
(zero crossings) occur earlier with respect to
 the preceding activity minima (indicated by the dotted vertical lines.)}
\label{DemonstrationEtar}
\end{figure*}

We use the results obtained in Sect.~\ref{sec:decaymodes} to extend the
surface transport model \citep{1985AuJPh..38..999D,WNS89} by a decay
term describing the volume diffusion of the poloidal magnetic field in
the convection zone. This is certainly only a rough description of
the evolution of the subsurface field, which can be improved once we
have more quantitative information about the systematic flows in the
deep convection zone. The working of the solar dynamo, on the other
hand, is already represented in the model by the emergence of bipolar
magnetic regions.

The extended equation for the evolution of the surface flux 
is written as
\begin{eqnarray}
\frac{\partial B_r}{\partial t} &=& -\omega(\theta)\frac{\partial B_r}{\partial \phi}  - \frac{1}{R_{\odot}\sin\theta}\frac{\partial }{\partial \theta} \bigg( v(\theta) B_r \sin\theta \bigg) \nonumber\\ 
\noalign{\vspace{0.2cm}}
&+& \frac{\eta_{\rm h}}{\Rsolar^2}\left[\frac{1}{\sin\theta}\frac{\partial }{\partial \theta} \bigg( \sin\theta\frac{\partial B_r}{\partial \theta}\bigg) + \frac{1}{\sin^2\theta}\frac{\partial^2 B_r}{\partial \phi^2} \right]
\nonumber\\ 
\noalign{\vspace{0.3cm}}
 &+& Q(\theta,\phi,t) - D(\eta)\,,
\label{extfluxtransportmodel}
\end{eqnarray}
where $\omega(\theta)$ is the angular velocity of the photospheric
plasma, $v(\theta)$ is the meridional flow velocity on the solar
surface, $Q(\theta,\phi,t)$ is a source term describing the emergence of
new magnetic flux, and $\eta_{\rm h}$ is the turbulent magnetic
diffusivity associated with the nonstationary supergranular motions on
the surface.  We specify the decay term $D(\eta)$ on the basis of the
decay modes in a spherical shell as determined in the previous section.
To this end, we expand the instantaneous radial surface magnetic field
into spherical harmonics,
\begin{eqnarray}
B_r \left(\Rsolar,\theta,\phi, t \right) = \sum_{l=1}^{\infty} 
   \sum_{m=-l}^{m=+l} c_{lm}(t) \;Y_{lm} \left(\theta,\phi \right) \, .
\label{SphericalHarms}
\end{eqnarray}
The volume diffusion leads to exponential decay of each of these modes
with the corresponding decay times, $\tau_{ln}= 1/(\eta\, k_{ln}^2)$,
depending on the radial structure of the magnetic field. Since the
latter is unknown in the framework of the flux transport model, we only
consider the mode $n=0$. This is justified by the fact that all higher
modes ($n\ge 1$) decay much more rapidly (see Table~\ref{table1}), so
that they do not affect the long-term, large-scale behaviour of the
surface field that the flux transport models aim to describe. We
therefore write $D(\eta)$ as
\begin{eqnarray}
   D(\eta) = \sum_{l=1}^{\infty}  \sum_{m=-l}^{m=+l} 
   \frac{c_{lm}(t)}{\tau_{l0}} \;Y_{lm} \left(\theta,\phi \right)\, .
\label{DecayTerm}
\end{eqnarray}
As we have seen in Sect.~\ref{sec:decaymodes}, the decay times do not
depend on the azimuthal wave number, $m$, of the mode. The modal decay term
given by Eq.~ (\ref{DecayTerm}) is particularly simple to implement in
flux transport codes based upon an expansion into surface harmonics
\citep[e.g.,][]{BCP98, Mackay1, Baumann2004}.

The decay term, $D(\eta)$, depends on the turbulent volume diffusivity,
$\eta$, which will generally differ from the diffusivity $\eta_{\rm h}$
related to the flux transport by the horizontal surface motions of
supergranulation. Since the velocities and length scales of the dominant
convective motions in the deep convection zone, which are relevant for
$\eta$, are not well known, in the following sections we empirically
estimate the value of $\eta$ by comparing with observed properties of
the solar polar fields, namely the times of polar reversals and the
sustained polar reversals in the case of solar cycles of varying
strength.

\section{Cycles of varying strength}
\label{sec:cycle_var}

One of the main problems with the original formulation of the flux
transport model is the too long memory of the system. In the case of cycles
with varying strength this can lead to a secular drift of the polar
fields and a suppression of polar reversals. In the case of random
fluctuations of the cycle amplitude, it results in a random walk of the
polar fields superposed upon the cyclical variation. We illustrate this
effects and their elimination through the decay term by way of a couple
of examples.

\begin{figure*}
\centering
\resizebox{0.8\hsize}{!}{\includegraphics[angle=270]{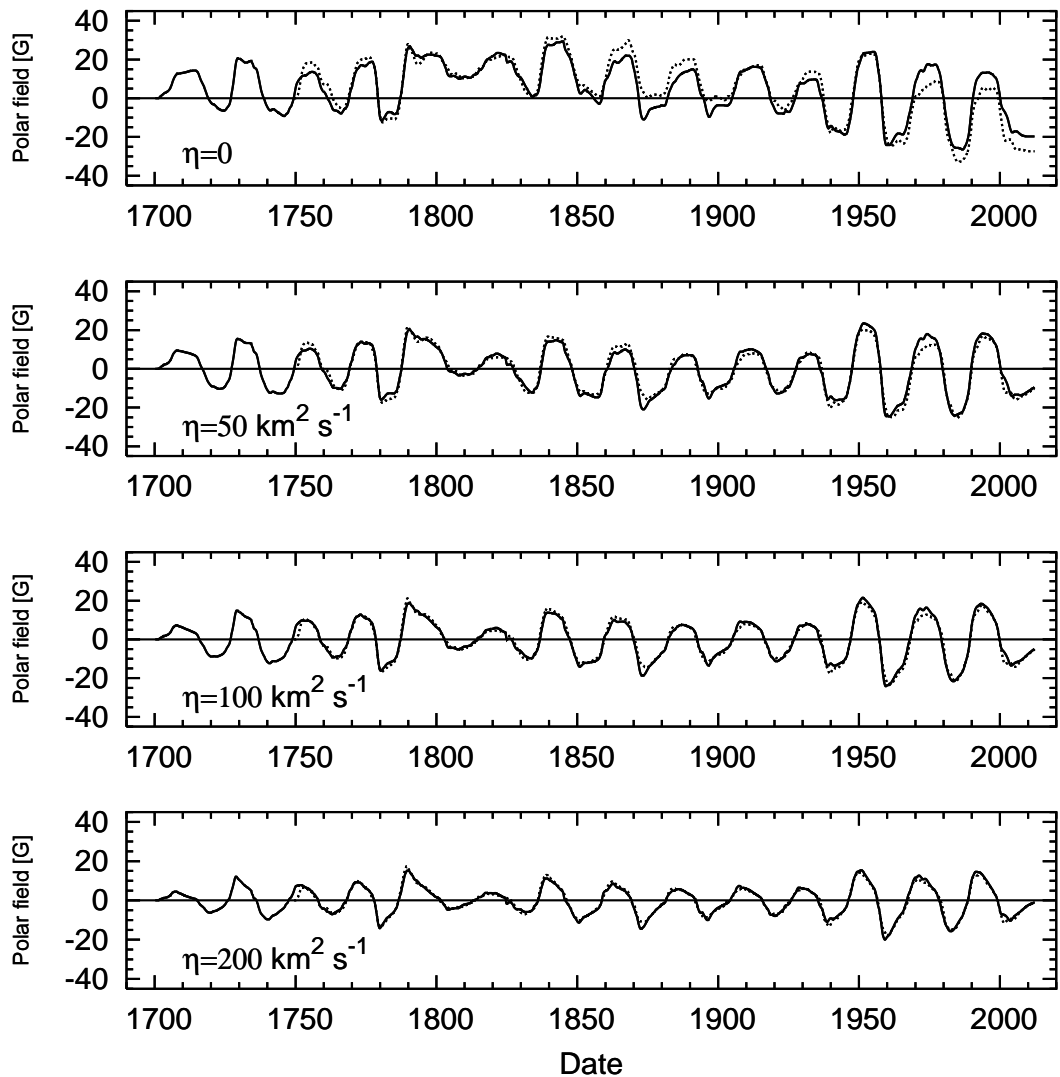}}
\caption{Evolution of the north polar field (average above $75^\circ$
latitude) for a sequence of synthetic cycles with the flux emergence
rate taken proportional to the sunspot numbers since 1700 (full line)
and 1750 (dotted line), respectively, for different values of the volume
diffusivity, $\eta$. The top panel ($\eta$=0) shows the strong drift of
the polar field resulting from the secular increase of solar activity
during the last century \citep[see also][]{sch01jan}. Finite values of
$\eta$ reduce the secular effect and lead to more symmetric variations,
as observed;  the volume diffusion term also largely removes the
dependence of the results on the initial conditions (difference between
full and dotted lines).}
\label{fig:cycles}
\end{figure*}

As a first illustration of the effect of the decay term introduced in
the flux transport model, we consider simulations of synthetic sets of
activity cycles \citep[see][for a detailed description of the
model]{Baumann2004} that vary systematically in strength: the cycle
amplitudes (total amount of emergent flux) alternate by a factor of 2 between
odd and even cycles.  The results for the polar fields for several
values of $\eta$ are shown in Fig.~\ref{DemonstrationEtar}. We define
the polar field, $B_{\rm pole}$, as the averaged radial field strength
over the polar cap poleward of $75 \dgr$ latitude. For $\eta=0$, i.e.,
in the absence of the added decay term, the polar fields show a strong
secular drift (Fig.~\ref{DemonstrationEtar}a). This results from the
fact that, during the weaker cycles, the amount of opposite-polarity
flux reaching the poles is insufficient to cancel the existing polar
field and to build up a field of opposite polarity. This results in a
systematic drift of the polar fields. Including the decay term leads to
a shorter memory of the system and a stabilization of the
oscillation (with some fluctuations due to the random component in the
prescription of the flux emergence). For growing values of $\eta$, the
drift of the polar fields ceases earlier and the asymmetry as well as
the amplitude of the oscillation decreases
(Fig.~\ref{DemonstrationEtar}b--d). At the same time, the sign reversals
of the polar fields occur earlier after the minima of flux eruption
(activity minima, indicated by the dotted vertical lines) because the
polar fields from the previous cycle have already been reduced by the
effect of the decay term.

We have also considered the secular variation of solar activity in the
historical record of sunspot numbers \citep[see also][]{sch01jan}. Using
the record of sunspot numbers since 1700, we have simulated a
series of solar cycles with the flux transport code. We have used the
`standard' parameters (butterfly diagram of emerging bipolar regions,
tilt angles, polarity rules, transport parameters, etc.) of
\citet{Baumann2004}, except for taking the emergence rate of the bipolar
regions proportional to the sunspot numbers (monthly sunspot
numbers since 1750 and monthly values interpolated from the yearly
sunspot numbers before), thereby also using cycle lengths in agreement
with the observed record. The overall free scaling factor for the flux
emergence has been fixed by requiring that the total (unsigned) surface
flux during the last three solar maxima (cycles 21--23) for the case
$\eta=100\,\kms$ matches the observed values given by
\citet{Arge:etal:2002}. The evolution of the polar fields according to
simulations with different values of $\eta$ is shown in
Fig.~\ref{fig:cycles}.  In order to evaluate the effect of the
arbitrary initial condition (zero surface field) we also show a run
starting from 1750, represented by the dotted line.  In the case
$\eta=0$ (no volume diffusion), the long memory of the system leads to a
drift of the polar fields in the $20^{\rm th}$ century due to the
secular increase of solar activity, so that the oscillations become
very asymmetric, in striking contrast to the observed evolution of the
polar fields.  Finite values for $\eta$ of the order $100\,\kms$ lead to
more symmetric oscillations and a suppression of the unrealistic
drift. The amplitude of the simulated polar field in the last cycles for
a value of $\eta=100\,\kms$ is roughly consistent with the published
observational data, which indicate amplitudes for the field strength of
10--20~G
\citep[e.g.,][]{Wang:Sheeley:1995,Arge:etal:2002,Dikpati:etal:2004,
Durrant:etal:2004}. Comparing the two runs with different starting
times (full line and dotted line), we find that the long memory of the
system in the case $\eta=0$ leads to a significant difference between
these two runs. For $\eta\neq 0$, on the other hand, there is almost no
dependence on the initial condition after a few cycles.

\section{Polar field reversal times}
\label{sec:reversals}

\begin{table*}
  \caption{Epochs of reversals (average of north and south pole) of
the simulated polar magnetic field poleward of $\pm75^\circ$ latitude
for three values of the magnetic volume diffusivity, $\eta$ (in $\kms$),
in comparison with the reversal times inferred by \citet{Makarov2003} on
the basis of polar crown filaments, $T_{\rm c}$, and from H$\alpha$
synoptic charts, $T_{\rm l=1}$. The row at the bottom gives the average
time intervals (in years) between the reversals and the preceding sunspot
minima ($T_{\rm min}$, last column).}
\centering
\begin{tabular}{cc|cccccc} \hline\hline 
& Cycle \# & $\eta=50$ & $\eta=100$ &
$\eta=200$ & $T_{\rm c}$ &  $T_{\rm l=1}$ & $T_{\rm min}$ \\ \hline 
& 13 & 1895.0 & 1893.7 & 1892.7 & 1895.0 & 1893.2 & 1889.6 \\ 
& 14 & 1907.8 & 1906.7 & 1905.2 & 1908.7 & 1905.8 & 1901.7 \\ 
& 15 & 1918.4 & 1917.5 & 1916.4 & 1918.7 & 1916.3 & 1913.6 \\ 
& 16 & 1928.9 & 1927.7 & 1926.8 & 1929.9 & 1927.0 & 1923.6 \\
& 17 & 1938.2 & 1937.7 & 1936.8 & 1940.1 & 1936.5 & 1933.8 \\ 
& 18 & 1949.5 & 1948.4 & 1947.3 & 1950.2 & 1947.3 & 1944.2 \\ 
& 19 & 1958.6 & 1957.9 & 1957.1 & 1959.7 & 1957.2 & 1954.3 \\ 
& 20 & 1971.6 & 1969.6 & 1968.3 & 1971.5 & 1968.6 & 1964.9 \\ 
& 21 & 1981.2 & 1980.6 & 1979.6 & 1981.8 & 1979.9 & 1976.5 \\ 
& 22 & 1991.3 & 1990.6 & 1989.7 & 1991.8 & 1990.4 & 1986.8 \\ 
& 23 & 2002.3 & 2001.3 & 1999.8 & 2001.7 & 1999.7 & 1996.4 \\ 
\hline 
& $\langle T_{\rm rev} - T_{\rm min}\rangle$ &
$5.2\pm0.8$ & $4.2\pm0.5$ & $3.1\pm0.2$ & $5.8\pm0.6$ & $3.3\pm0.5$ & \\
\hline
\end{tabular}
\label{tab:reversal}
\end{table*}

Having shown that the decay term prohibits drifts and an unrealistic
long-term memory of the polar fields for values of $\eta$ of the order
of $100\,\kms$, we now consider its effect on the calculated reversal
times of the polar fields and compare quantitatively with observational
results.

\begin{figure}
\centering
\resizebox{\hsize}{!}{\includegraphics{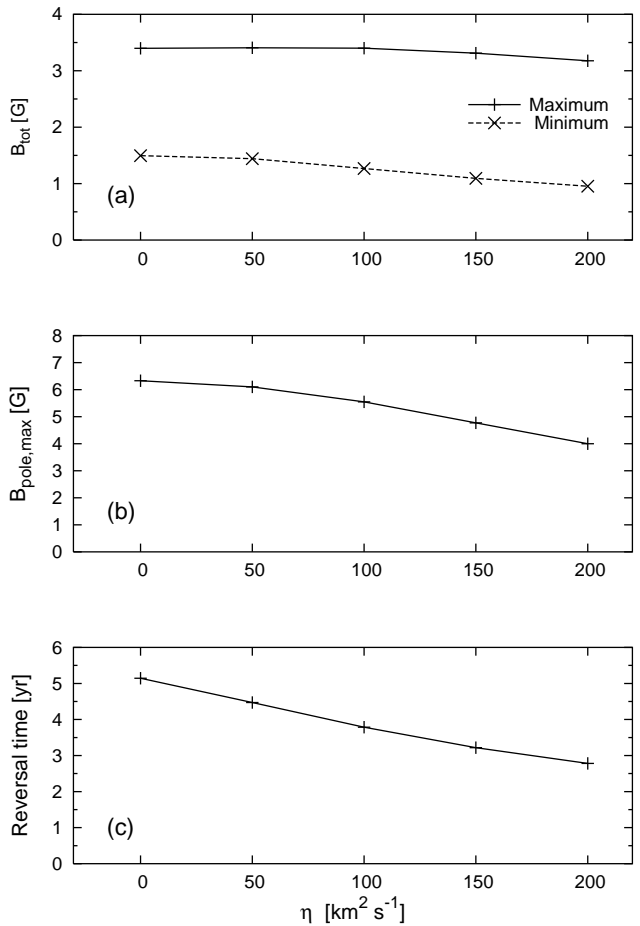}}
\caption{Synthetic solar cycles: dependence on $\eta$ of the (a)
averaged unsigned surface field (maxima and minima, respectively), (b)
the maxima of the polar polar field, and (c) the polar reversal times
after the previous activity minimum (minimum of the cyclic flux
emergence rate in the model).}
\label{crosscor}
\end{figure}

\subsection{Synthetic cycles}
\label{subsec:rev_synthetic}

In order to illustrate the effect of the decay term on the reversal
times, we consider synthetic cycles of equal strength and determine the
reversal times of the polar field in dependence on the value of
$\eta$. In addition, we also consider the averaged (unsigned) field over
the whole surface, $B_{\rm tot}$, and the maximum polar field during a
cycle. The absolute values of the field strength are arbitrary; here we
are only concerned with the dependence on $\eta$.  Fig.~\ref{crosscor}
shows: $B_{\rm tot}$ for cycle maxima and minima, respectively
(Fig.~\ref{crosscor}a), the maximum polar field (Fig.~\ref{crosscor}b),
and the reversal time in years after the previous flux emergence
(activity) minimum (Fig.~\ref{crosscor}c), all as functions of $\eta$.

While the average surface field during cycle maxima varies only slightly
with $\eta$, there is a somewhat stronger decline of the values during
cycle minima (Fig.~\ref{crosscor}a), similar in proportion to the
decline of the maximum polar field with increasing $\eta$
(Fig.~\ref{crosscor}b). This results from the fact that, during
activity minimum, both the polar and the total field are dominated by
the dipole mode, which suffers from enhanced decay due to volume
diffusion.  A strong effect of the value of $\eta$ on the polar reversal
times is clearly visible in the lower panel. For increasing $\eta$,
the reversals occur earlier after cycle minimum, varying between 5.1
years for $\eta=0$ and 2.8 years for $\eta=200\kms$. This results from
the stronger decay of the polar field from the previous cycle, so that
less opposite-polarity flux of the new cycle is required to reverse the
field, leading to an earlier reversal time.

Although these synthetic cycles are not intended for a quantitative
comparison with actual solar data since they do not include a long-term
modulation of the cycle amplitudes and also no variation of the cycle
length, we may nevertheless check whether they are roughly consistent
with the data. \citet{Makarov2003} used two different methods to
determine polar reversal times for the period 1878--2001: (1) by the
disappearance of the polar crown filaments, and (2) from estimating the
magnetic dipole configuration on the basis of H$\alpha$ synoptic charts.
They find that reversals occur on average 5.8 years after the previous
sunspot minimum for method (1) while they obtained a value of 3.3 years
with method (2). \citet{Makarov2003} consider the first method to be
more reliable. On the other hand, the magnetograph data compiled by
\citet{Arge:etal:2002} and \citet{Dikpati:etal:2004} for the the solar
cycles 21--23 indicate an average reversal time about 4.6 years after
sunspot minimum, about the mean of the two values given by
\citet{Makarov2003}. Comparing this with the bottom panel of
Fig.~\ref{crosscor}, we find that a value of $\eta$ in the range
50--100$\,\kms$ leads to reversal times that are consistent with
these results. A quantitatively more reliable comparison is carried out in
the subsequent section.

\subsection{Simulation of solar cycles No. 13 -- 23 on the basis of
            RGO/SOON sunspot data}
\label{subsec:solar_cycles}

For a quantitative comparison with observations, we determine the
magnetic flux input into the flux transport model on the basis of the
sunspot group areas from the digitized version of the Royal Greenwich
Observatory (RGO) photographic results, which are available for the time
period 1874-1976. For the time after 1976, this series is continued by
data from the Solar Optical Observing Network (SOON) of the US Air
Force.  We have combined both datasets and transformed them into a
sequence of emerging active regions. Below we give a brief summary
of the procedure; a more detailed description will be given elsewhere.

We take each observed sunspot group to provide a bipolar magnetic region
and determine its total unsigned magnetic flux from the observed sunspot
area and an associated facular (plage) area according to the empirical
relationship derived by \citet{CCD97}. We use the sunspot groups in the
databases at the time of their maximum area development, thus ensuring
that every group is considered only once.  Furthermore, we take only
sunspot groups that reach their maximum area within $\pm 45^{\circ}$ of
the central meridian. Owing to the large number of observations, we
consider this $90^{\circ}$ window to be representative for the whole
sun.  In order to roughly account for the flux emergence on the full
solar surface, we copy this $90^{\circ}$ window three times into the
remaining $270^{\circ}$ longitude range.  Each of the sunspot groups
determined in this way provides a bipolar magnetic region whose
orientation of the magnetic polarities is assigned according to Hales'
polarity rules and whose tilt angle, $\alpha$, is assumed to be given by
Joy's law, i.e. $\alpha=0.5\,\lambda$, where $\lambda$ is the
latitude. Further details about the treatment of bipolar magnetic region
are given in \citet{Baumann2004}. The overall scaling factor in the
magnetic flux of the emerging bipolar regions is chosen such that the
flux transport simulations reproduce the evolution of the observed total
surface field given by \citet{Arge:etal:2002} and the polar field
strengths given by \citet{Dikpati:etal:2004}.

\begin{figure*}
\centering
\resizebox{0.9\hsize}{!}{\includegraphics{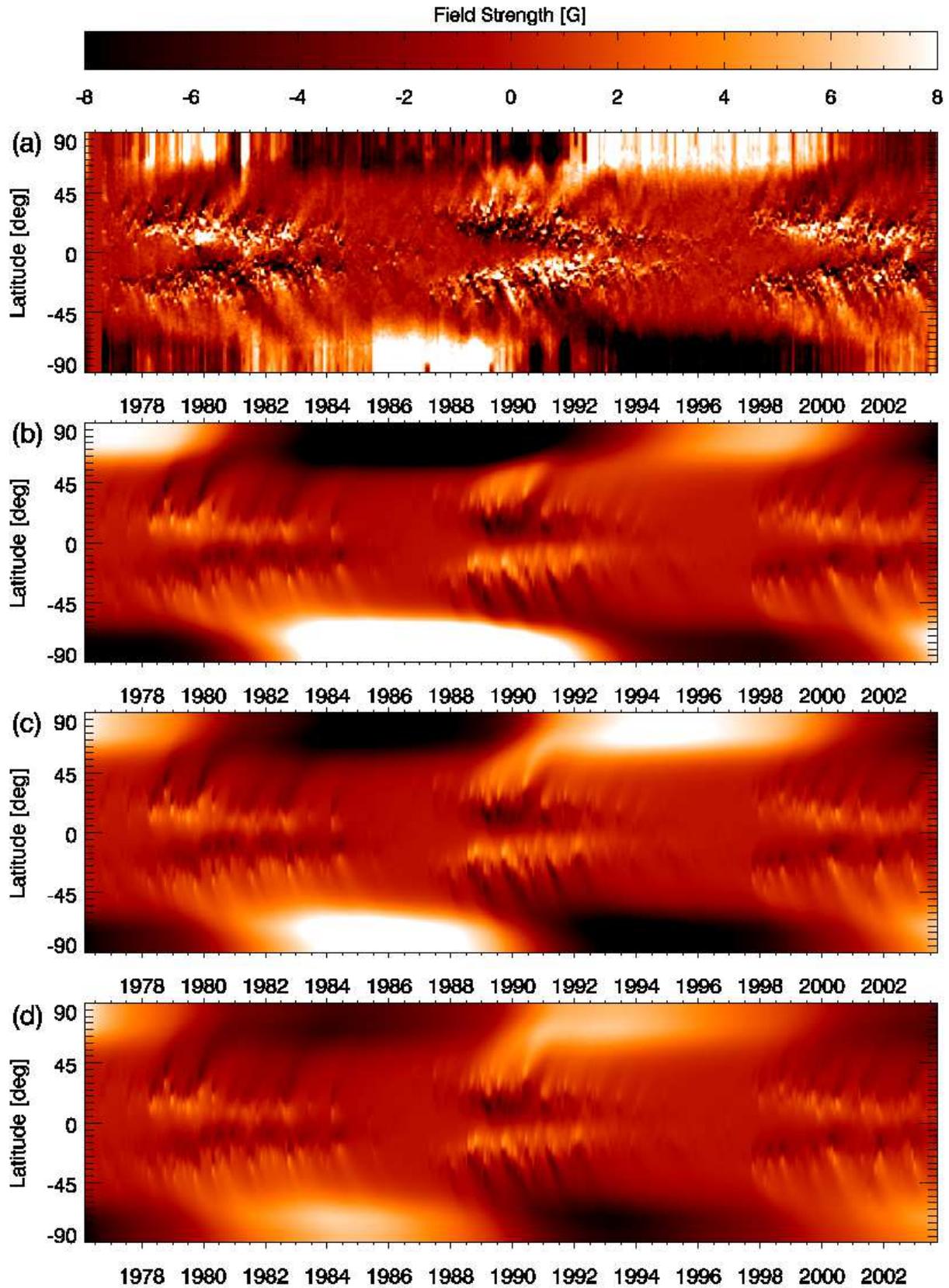}}
\caption{Comparison between observed magnetic field distribution
from 1974 on (panel a, based upon NSO/Kitt Peak data, courtesy
D. Hathaway) and flux transport simulations for different values of the
volume diffusivity: $\eta=0$ (panel b), $\eta=100\kms$ (panel c), and
$\eta=200\kms$ (panel d). Shown are time-latitude diagrams of the
longitudinally averaged surface field.  The simulations reproduce the
global development of the magnetic field, with poleward surges of
following-polarity flux leading to reversal of the polar fields around
sunspot maxima. However, in the case without the decay term (panel b),
the reversals occur too late. For non-vanishing values of $\eta$, the
reversal times are shifted forward in time; the best agreement with the
observed reversal times is obtained for $\eta=100\kms$ (panel c). }
\label{herringbone}
\end{figure*}

We have carried out a flux transport simulation based upon the set of
input data covering the period 1874--2005. For the interval 1974--2005,
the time-latitude diagrams of the longitudinally averaged magnetic field
from the simulation can be compared with the corresponding diagram
derived by D. Hathaway (NASA/Marshall Space Flight Center) from the
NSO/Kitt Peak synoptic maps. Fig.~\ref{herringbone} shows this
comparison for three values of the volume diffusivity:
$\eta=0,\,100,\,200\kms$. The evolution of the global field is best
reproduced by the case $\eta=100\kms$, which yields the nearest
agreement of the reversal times of the polar fields for the last three
cycles.

In order to compare with the result of \citet{Makarov2003}, we have
determined the epochs of the reversals of the average field poleward of
$\pm75^\circ$ latitude from the flux transport simulations for the solar
cycles 13--23.  The reversal times (average of the north and south polar
reversals) are given in Table~\ref{tab:reversal} together with the
reversal times inferred by \citet{Makarov2003} from the evolution of
polar-crown filaments ($T_{\rm c}$) and from H$\alpha$ synoptic charts
($T_{\rm l=1}$), respectively. The last row gives the average time
interval (in years) between the polar reversal and the preceding solar
minimum. We recover the trend towards earlier reversals for larger
values of $\eta$ found in the previous section for the synthetic
cycles. The values cover about the range defined by the results from the
two methods used by \citet{Makarov2003}. However, even with
$\eta=50\kms$ we do not reach the value of 5.8 years that they
determined from the disappearance of polar crown filaments. This could
be due to the fact that we define the polar reversal from the {\it
average} field in the caps poleward of $\pm 75^\circ$ latitude, while
the polar crown filaments probably disappear later, when the last remnant
of the old polarity in fact vanishes. A fair degree of averaging is
also inherent in the magnetograph data shown by
\citet{Dikpati:etal:2004}, from which we roughly estimate an average
reversal time of 4.6 years after sunspot minimum. Altogether, we
conclude that a volume diffusivity in the range 50--100 $\kms$ is
probably adequate for flux transport models of the solar surface
field. This is also consistent with the timings of the polar reversals
with respect to the sunspot maxima: for $\eta=50\kms$ the polar field
reverse on average 1.3 years after solar maximum; for $\eta=100\kms$ we
find an average value of 0.3 years after maximum.

\section{Conclusion}
\label{sec:conclusion}

We have shown that a modified (modal) version of the decay term first
introduced into the flux transport model by \citet{sch02oct} can be
derived consistently from the volume diffusion process that is (among
other factors) neglected in the standard flux transport
models. Including this term removes the unrealistic long memory of the
system and thus prohibits a secular drift or a random walk of the polar
fields in the case of activity cycles with variable amplitude.  The
value of the (turbulent) magnetic volume diffusivity, $\eta$, can be
estimated by considering the reversal times of the polar field relative
to the preceding activity minimum and by comparing with direct
observations or values inferred from proxy data.  We find that values of
$\eta$ in the range 50--100$\,\kms$ are consistent with the observational
constraints. This is also the order of magnitude suggested by simple
estimates based on mixing-length models of the convection zone. With
this consistent extension and improvement of the model, flux transport
simulations of the large-scale magnetic field on the solar surface over
many activity cycles can be carried out.

\acknowledgements{We are grateful for very helpful comments by an
anonymous referee. David Hathaway (NASA/Marshall Space Flight Center)
kindly provided the data for Fig.~\ref{herringbone}a.}

\bibliographystyle{aa} 
\bibliography{3488.bbl}

\end{document}